Cover Note (NCLIM-20040830C)

**Title: Carbon loss from forest degradation exceeds that from deforestation in the Brazilian Amazon**

**The length of the text, methods and legends:**

Abstract: 150 words

Main text: 3400 words

Methods: 3022 words

Legends: 544 words

**The number of references:**

Main text: 51 references

Methods: 18 references

**The number and estimated final size of figures and tables:**

5 figures (5 MB) in the Main Text section

**Title**:

**Carbon loss from forest degradation exceeds that from deforestation in the Brazilian Amazon**

**Author list**:

Yuanwei Qin[1], Xiangming Xiao[1,*], Jean-Pierre Wigneron[2,*], Philippe Ciais[3], Martin Brandt[4], Lei Fan[5,2], Xiaojun Li[2], Sean Crowell[6], Xiaocui Wu[1], Russell Doughty[1,7], Yao Zhang[8], Fang Liu[9], Stephen Sitch[10], Berrien Moore III[6]

**Affiliations**:

[1]Department of Microbiology and Plant Biology, Center for Earth Observation and Modeling, University of Oklahoma, Norman, OK 73019, USA

[2]ISPA, UMR 1391, INRA Nouvelle-Aquitaine, Bordeaux Villenave d'Ornon, France

[3]Laboratoire des Sciences du Climat et de l'Environnement, LSCE/IPSL, CEA-CNRS-UVSQ, Université Paris-Saclay, 91191 Gif-sur-Yvette, France

[4]Department of Geosciences and Natural Resource Management, University of Copenhagen, Copenhagen, Denmark

[5]Chongqing Jinfo Mountain Karst Ecosystem National Observation and Research Station， School of Geographical Sciences, Southwest University, Chongqing 400715, China

[6]College of Atmospheric and Geographic Sciences, University of Oklahoma, Norman, OK, 73019, USA

[7]Division of Geological and Planetary Sciences, California Institute of Technology, Pasadena, CA, USA

[8]Department of Earth and Environmental Engineering, Columbia University, New York, NY 10027, USA.

[9]Institute of Geographic Sciences and Natural Resources Research, Chinese Academy of Sciences, Beijing 100101, China

[10]College of Life and Environmental Sciences, University of Exeter, Exeter EX4 4RJ, UK

**Corresponding Author**:

[*]To whom correspondence should be addressed. Email: xiangming.xiao@ou.edu; jean-pierre.wigneron@inrae.fr

**Abstract**

**Spatial-temporal dynamics of aboveground biomass (AGB) and forest area affect the carbon cycle, climate, and biodiversity in the Brazilian Amazon. Here we investigate inter-annual changes of AGB and forest area by analyzing satellite-based annual AGB and forest area datasets. We found the gross forest area loss was larger in 2019 than in 2015, possibly due to recent loosening of forest protection policies. However, net AGB loss was three times smaller in 2019 than in 2015. During 2010-2019, the Brazilian Amazon had a cumulative gross loss of 4.45 Pg C against a gross gain of 3.78 Pg C, resulting in net AGB loss of 0.67 Pg C. Forest degradation (73%) contributed three times more to the gross AGB loss than deforestation (27%), given that the areal extent of degradation exceeds deforestation. This indicates that forest degradation has become the largest process driving carbon loss and should become a higher policy priority.**

**Main text**

Tropical forests in the Amazon account for approximately 50% of the rainforests in the world[1] and are important for global biodiversity, hydrology, climate, and the carbon cycle[2,3,4,5,6]. Accurate and timely data on vegetation aboveground biomass (AGB) and forest area in the region at various spatial and temporal scales are needed to understand the carbon balance, which is affected by land use, logging and degradation, secondary forest regrowth, and climate[7,8]. In addition to *in-situ* AGB measurements in intact forests[9,10,11], several studies combined *in-situ* AGB data with images from optical, microwave and laser sensors to generate static AGB maps over merged periods (e.g., circa 2000[3], circa 2007-2008[12], and 2003-2014[13]). Combined with forest area change datasets from the Amazon Deforestation Monitoring Project (PRODES)[14] and the Global Forest Watch (GFW)[15], these static AGB maps are used to estimate AGB dynamics from deforestation and forest degradation[13,16], but forest losses from PRODES were substantially smaller than from GFW[17,18,19]. These differences and uncertainties result from different forest definitions and the use of Landsat images, which are severely impacted by frequent clouds and aerosols from fire, leaving only very few good-quality images per year[17]. This issue could be solved by the use of Moderate Resolution Imaging Spectroradiometer (MODIS) data, which spatial resolution cannot identify small patches of forest losses or gains, but the daily images ensures more good-quality observations per year[17].

Substantial progress has been made in analyzing L-band Vegetation Optical Depth (L-VOD) from the Soil Moisture and Ocean Salinity (SMOS) passive microwave images, which provides annual maps of AGB since 2010 at 0.25° spatial resolution (see Methods)[20, 21, 22, 23]. Moreover, images from the Phased Array type L-band synthetic Aperture Radar (PALSAR) and MODIS were used to derive annual maps of evergreen forest areas at 500-m resolution for the Brazilian Amazon during 2000-2017[17, 24]. Combining L-VOD AGB and PALSAR/MODIS forest area during 2010-2019 offers a unique window to assess the spatial-temporal dynamics of AGB and forest area in the Brazilian Amazon, and how this is impacted by climate and land use. This period is of special interest, because the impacts on forest area and biomass from extreme climate events and the changed policies of the new Brazilian government (in office since January 2019), favoring the expansion of pasture[25, 26] at the expense of forest conservation, has not yet been fully quantified.

Here we used annual L-VOD AGB[20] and annual forest area datasets[17] described above to investigate the spatial-temporal dynamics of forest carbon in the Brazilian Amazon during 2010-2019. We investigated (1) the role of climate anomalies in the changes of forest area and AGB, e.g., Atlantic Multi-decadal Oscillation (AMO, 2010), El Niño (2009/2010 and 2015/2016), La Niña (2010/2011 and 2017) (Extended Data Fig. 1); (2) whether recent changes of policies and human activities in 2019 have a detectable effect on forest area and AGB; and (3) the relative contribution of deforestation and forest degradation (forest fragmentation, edge-effects, logging, forest fire and drought) to inter-annual variation of AGB loss in the study period.

**Consistency between AGB and forest area**

The AGB and forest area data were organized into 5,656 grid cells at 0.25° spatial resolution (longitude, latitude; ~25-km × 25-km) (Methods). We studied the relationships between annual AGB and forest area fraction (%, FAF) for individual grid cells. The spatial distribution of AGB agrees well with that of FAF in 2019 (Fig. 1a,b). AGB and FAF are linearly (spatially) correlated with each other in 2019 and other years (Fig. 1c, Extended Data Fig. 2, $R^2 \geq$ 0.81). We also investigated the temporal consistency between AGB and FAF for all grid cells over the ten years. As an example, we showed two contrasting grid cells that exhibited either a large loss (Fig. 2a,b,c) or a large gain (Fig. 2d,e,f) in FAF. The temporal correlation between AGB and FAF (AGB = f (FAF)) was found to be strong in the 'loss' grid cell ($R^2 = 0.82$, $P$-value < 0.01) and lower but significant for the 'gain' grid cell ($R^2 = 0.30$, $P$-value < 0.1). The spatial distributions of the temporal relationships between AGB and FAF during 2010-2019 are shown in Fig. 2g,h. We found that 23% of the total area ($112 \times 10^6$ ha) had a statistically significant ($P$-value < 0.05) and positive linear relationship between AGB and FAF, especially in the southern and eastern Brazilian Amazon. This loss of AGB following forest area losses is expected, but the slope of the relationship differs depending on the mechanisms that lead to forest area loss and exposed AGB densities. On the other hand, intact forest with no forest area loss can have changes in AGB due to climate anomalies or forest degradation. At 25-km spatial resolution, we only observed the bulk of AGB changes and further work is needed to attribute the roles of forest area loss, forest area gain, as well as forest degradation on top of climate-induced variability. In the following, we take a closer look at yearly anomalies to gain insights on those drivers.

**Inter-annual changes of AGB and forest area**

The inter-annual changes of forest area, active fire area, burned area, and AGB are displayed in Fig. 3. We decomposed annual net AGB change into the sum of gross AGB loss (grid cells with negative changes) and gross gain (grid cells with positive changes). The gross forest area loss in 2019 (3.9× 10$^6$ ha), which was a drought year, was larger than that during the extreme El Niño drought year of 2015 (3.0 × 10$^6$ ha) (Fig. 3c). This suggested that the combined impacts of policy changes by the Brazilian government[25, 26] and drought (i.e., drought-induced tree mortality and enhanced forest fires) caused a larger forest area loss in 2019. In contrast, the net AGB change in 2019 (-0.05 Pg C) was only one fifth of the net AGB change in 2015 (-0.25 Pg C) (Fig. 3d), which is confirmed by the large gross AGB loss (-0.55 Pg C) in 2015 (Fig. 3d, Extended Data Fig. 3). The strong El Niño in 2015 thus resulted in a more extensive loss of AGB over both intact and secondary forests, from drought and drought-induced fires coordinated with human ignitions[27]. Comparing losses of AGB and forest area changes in 2014/2015 and 2018/2019 revealed that the extreme El Niño in 2015 and the combined impact of policy changes and drought in 2019 had a differential impact on AGB and forest area. As 2019 was the first year of Brazilian President Jair Bolsonaro's administration, the impacts of those policy changes on AGB and forest area remain to be investigated beyond 2020.

Over the 10-year period, linear regression analysis showed a strong correlation between annual AGB and forest area (Fig. 3a, $R^2$ = 0.78). Annual AGB decreased from 44.86 Pg C in 2010 to 44.19 Pg C in 2019, a net loss of 0.67 Pg C (0.07 Pg C yr$^{-1}$), while annual forest area decreased from 370.21 × 10$^6$ ha in 2010 to 361.29 × 10$^6$ ha in 2019, a net loss of 8.91 × 10$^6$ ha (0.99 × 10$^6$ ha yr$^{-1}$) (Fig. 3a). These total numbers mask the highly dynamic and regional nature of inter-annual changes in gross gains and gross losses of AGB and forest area, which partly compensate for each other. We thus calculated inter-annual changes of AGB (Fig. 3d) and forest area (Fig. 3c) between two consecutive years for individual grid cells, and identified gross gains and gross losses as the sums of AGB-changes on all the grid cells showing either gains or losses (Fig. 3e). On average, gross AGB loss and gain (Fig. 3e) were about five times larger than net changes between two years (Fig. 3d). The cumulated gross loss and gross gain of AGB in the Brazilian Amazon over 2010-2019 were 4.45 Pg C and 3.78 Pg C, respectively.

The cumulated gross forest area loss over the 10 years was about $19.75 \times 10^6$ ha (Fig. 3e). The GFW[18] reported loss ($19.14 \times 10^6$ ha during 2010-2018) is very close to our estimate. PRODES reported only $6.72 \times 10^6$ ha forest area loss during 2010-2019[14], which is because it was designed to monitor only deforestation of old-growth primary forests as per 1988, not considering losses from secondary forests which have a high turnover and can get deforested several times within our study period[28]. The GFW and our MODIS forest area datasets include losses of primary and secondary forests since 2000 and 2001, respectively. From 1988 to 2000, some pixels classified as intact forest in 1988 by PRODES may have already been deforested and regenerated when their dynamics is monitored by MODIS and GFW products.

We calculated the temporal dynamics of AGB and FAF for six classes of FAF (Methods) and found that AGB varied temporally in tandem with FAF (Extended Data Fig. 4), suggesting that inter-annual changes in forest area are one of the major factors contributing to inter-annual changes of AGB. The inter-annual variations of active fire and burned areas (Fig. 3b) corresponded well with those of annual AGB and forest area losses during 2010-2019, except for 2017 and 2019 (Fig. 3c,d), indicating that fire was strongly associated with the losses of AGB and forest area.

**AGB and forest area losses in El Niño years**

The impacts of El Niño climate events on vegetation have been debated intensively over the past decades[10, 11, 27, 29, 30]. Seasonally-moist Amazonian forests have deep root systems that could use water in deep soils, and they have a relatively high resilience to drought[11, 31]. We calculated inter-annual changes of AGB and forest area between the 2015 extreme El Niño year and the previous year (Extended Data Fig. 1a). The net AGB change was negative and the largest in 2015 (-0.25 Pg C), with a gross AGB loss of 0.55 Pg C that surpassed a modest gross AGB gain of 0.29 Pg C (Fig. 3d). Net forest area change (-5.79 $\times 10^6$ ha) was also large in 2015. We detected a much larger loss of forest area in 2015 than in 2016, but the GFW and PRODES datasets showed smaller losses of forest area in 2015 than in 2016 (Extended Data Fig. 5). This discrepancy can be attributed to different definitions of forest[16], mapping algorithms (PRODES excluding secondary forest loss), calendar year (PRODES using August current year to July subsequent year), and the limited number of Landsat images used by the GFW and PRODES projects. The larger loss of AGB (Fig. 3a), larger active fire area and burned area (Fig. 3b), and larger annual growth rate of atmospheric $CO_2$ concentration (Fig. 3f, Extended Data Fig. 6) in 2015 supports our finding of a larger loss of forest area in 2015 than in 2016.

To identify hotspots of AGB and forest area change in 2015, we calculated the changes of average AGB and forest area during 2010-2013 and during 2015-2018 (Fig. 4a,b). The spatial distribution of AGB change (Fig. 4a) matched well with that of forest area change in the "Arc of Deforestation" (Fig. 4b). Between these two periods, AGB gain occurred in 29.40% of the area (141.71× $10^6$ ha) and AGB loss in 70.60% (340.26× $10^6$ ha) (Fig. 4a). Forest area gain occurred in 15.43% of the area (74.39× $10^6$ ha) and forest area loss in 51.63% (248.82× $10^6$ ha) (Fig. 4b). A fraction of 44.78% of the Brazilian Amazon had both AGB and forest area loss (Fig. 4c). The relationship between AGB and forest area changes between these two periods was statistically significant ($P$-value < 0.01) but weakly correlated (Fig. 4c). This partial decoupling between AGB and forest area happens because few grid cells have large losses of AGB and forest area, while many others have a small loss of forest area and moderate loss of AGB (Fig. 4d). These results show that in 2015, the contribution of deforestation to the AGB loss was moderate ($R^2$ = 0.19) and suggest that climate-induced tree mortality and degradation contributed to the AGB loss.

We further analyzed the inter-annual changes of AGB in those grid cells with stable forest area in relation to changes in mean annual precipitation and mean maximum cumulative water deficit (MCWD, Methods) in 2010-2013 and 2015-2018 (Fig. 4e, Supplementary Figs. 1, 2, Extended Data Fig. 7). Approximately 37% of this area (58.37× $10^6$ ha) had AGB gains (0.06

Pg C, 0.49 Mg C ha$^{-1}$ yr$^{-1}$), most of which was distributed in the northwest (Fig. 4a,b) where mean annual precipitation was higher than 2,000 mm yr$^{-1}$ (Supplementary Fig. 1). The remaining areas with no forest area change (63%; 99.53 × 10$^6$ ha) had AGB losses (0.14 Pg C; 0.70 Mg C ha$^{-1}$ yr$^{-1}$), suggesting that extensive forest degradation occurred. AGB loss increased as annual precipitation decreased, indicating that drought was a driver of forest degradation, and grid cells with an annual precipitation of < 2,000 mm yr$^{-1}$ had the largest sensitivity to drought (Fig. 4e). Approximately 85% of fires consistently occurred in the region with annual precipitation of < 2,000 mm yr$^{-1}$ and with a 70% increase of fires in the region with annual precipitation of ≥ 2,000 mm yr$^{-1}$, which can also explain AGB loss patterns in El Niño years[27]. In addition, for grid cells with forest area losses between the pre- and post- 2015 El Niño periods, AGB losses between these two periods were impacted by climate, deforestation, and human-induced forest degradation, which tended to be higher in areas with precipitation in the range 1500-2500 mm yr$^{-1}$ (Fig. 4f). Similar results are obtained with MCWD instead of mean annual rainfall (Extended Data Fig. 7). Further analyses of new land cover and degradation datasets[28, 32] is needed to attribute these bulk reductions of AGB to the different drivers and their interactions affecting forests in different regions of the Brazilian Amazon.

The 2015/2016 El Niño caused widespread AGB losses in 63% area of the Brazilian Amazon. We calculated the recovery strength[22] in the following three years and found that AGB fully recovered only in 25% of the area (Extended Data Fig. 8). The moist forest in the northwest and the Cerrado area in the southeast recovered quickly. In contrast, the "Arc of Deforestation" region where fires also peaked during the El Niño did not show a recovery of AGB. In this region, deforested areas from intact or secondary forests are primarily used for crops and pasture. Recent data show that secondary forests did regrow but were frequently deforested again[28].

**Increased AGB in La Niña years**

Several local studies investigated the speed of vegetation recovery in the Amazon after El Niños[10, 11, 33]. Our data with full coverage of the region show that forest area changed little between 2010 and 2012, but annual AGB in the strong La Niña of 2011 was higher (by 0.47 Pg C) than in the drought year 2010 (Fig. 3a,d). Field data from long-term forest plots reported slightly higher forest growth in 2011 as compared to 2010[10]. Results from atmospheric inversion suggested that in 2011 the Amazon Basin was a net $CO_2$ sink of $0.25 \pm 0.14$ Pg C yr$^{-1}$, higher than in 2010[34]. Similarly, annual AGB during the 2017 La Niña was also slightly higher (by 0.05 Pg C) than in the previous year, but this signal is mixed with the legacy effects of the 2015 El Niño (Fig. 3a). We also analyzed atmospheric $CO_2$ concentrations data over the Amazon and adjacent areas of the Atlantic and Pacific Ocean (from 10° N to 10° S) during 2015-2018 using $XCO_2$ data from NASA's Orbital Carbon Observatory (OCO-2) (Fig. 3f). The annual growth rate of $XCO_2$ over the Brazilian Amazon in 2016 (0.87 ppm), 2017 (1.80 ppm), and 2018 (1.79 ppm) were substantially lower than that in 2015 (3.51 ppm) (Fig. 3f, Extended Data Figs. 6, 9). This suggested that the $XCO_2$ gradient between the Amazon and surrounding oceans was more negative and was consistent with enhanced $CO_2$ uptake after the 2015 El Niño.

**AGB losses from both deforestation and degradation**

The loss of AGB observed in a 0.25° grid cell can be a mix of deforestation, reduction of biomass density from a suite of other processes, and a contribution from non-forest biomes, the latter with a smaller contribution to grid-cell AGB because of the low AGB of short vegetation. AGB decreases in the Brazilian Amazon have been attributed to direct human-induced deforestation, selective logging[35], forest fragmentation and associated edge effects[36], forest fires[27], and mortality from climatic disturbances like storms[37] and drought[38, 39]. Here we define forest degradation to include all these mechanisms that do not result in deforestation.

The contributions of deforestation and forest degradation to AGB losses cannot be explicitly separated within each 0.25° grid cell, but we performed a simple calculation based on a method reported by Harris et al.[40] (Methods, Fig. 5). Out of the cumulative gross AGB losses (4.45 Pg C) over the study period, we estimated that ~27% (1.18 Pg C) result from deforestation and ~73% (3.27 Pg C) from forest degradation, the latter being composed of 2.88 Pg C in grid cells with deforestation and 0.39 Pg C in the grid cells with no deforestation. Previous studies[9, 41] from local inventories and bookkeeping models[40] estimated that forest degradation contributed about 29%[9] or 18-40%[41] to the gross AGB losses in the Brazilian Amazon (Supplementary Table 1), which was less than our 'top-down' estimate of 0.25° L-VOD AGB loss. This can be explained by the full spatial coverage of the entire Brazilian Amazon, and because we included 'degradation' from climatic disturbances. Our result is in agreement with two previous studies[13, 42] (Supplementary Table 1). Aragão et al. presented a bottom-up carbon balance for the Brazilian Amazon decomposing each flux and separating drought effect, which showed that forest degradation contributed 65% to the AGB losses in 2000s[42]. Baccini et al. used Landsat-based forest cover data during 2003-2014 and estimated that forest degradation contributes 69% to the AGB losses in tropical forests[13]. Long-term forest degradation areas (337,427 km$^2$) surpassed deforestation (308,311 km$^2$) in the Brazilian Amazon during 1992-2014[32]. According to our estimate, AGB losses from forest degradation are substantial, and need to be explicitly included into the global carbon budget assessments[43]. Reducing forest degradation must be a policy priority in the Brazilian Amazon to reach the requirement of "Reducing Emissions from Deforestation and Forest Degradation" (REDD+) and the carbon emission reduction commitment of the 2015 Paris Agreement.

In areas of intact forests (defined with a > 99% persistent forest cover), AGB losses during 2010-2019 amounted to 0.10 Mg C ha$^{-1}$ yr$^{-1}$ and were found to be substantially associated with fire and water deficit (Extended Data Fig. 10). The AGB density change over intact forests was close to the average (0.06 Mg C ha$^{-1}$ yr$^{-1}$) estimated by the forest plots networks during 2000-2011[10]. During 2010-2015, intact forest AGB changes were highly temporally associated with water deficit ($R^2$ = 0.81, $P$-value < 0.01). During 2015-2019, although the water deficit was reduced, forest AGB continued to decrease due to the legacy effects of drought and a doubling of forest fires compared as to 2010-2014, which is supported by field measurements[44, 45, 46].

Forest conservation is a challenging task under severe droughts and governmental policies that threaten Amazon forests[47]. Here, we used two new satellite data products to quantify spatial-temporal changes of AGB and forest area in the Brazilian Amazon. The strong spatial-temporal consistency between annual AGB and FAF within individual grid cells during 2010-2019 enables us to attribute the relative contribution of deforestation and forest degradation to the losses in AGB[13, 48], potential carbon emissions to the atmosphere in a long period[44, 45, 46]. Large AGB losses in 2015–2016 and large AGB gains in 2011 and 2017 show that the forests are geographically divergent in their sensitivity and resilience to changes in climate, land use, and disturbance. Continued land use change[7, 26], increased climate extremes in the coming decades[38, 39], and new Brazilian governmental policies may reduce the capacity of the forests to sequester carbon[10, 11] and make it more challenging to achieve the objectives of the REDD+ program. To effectively manage, conserve, and monitor tropical forests, it is essential to fully integrate *in-situ*, citizen science, aerial, and space-borne data. Recently launched and future spaceborne platforms that measure characteristics of vegetation canopy and structure (Global Ecosystem Dynamics Investigation (GEDI)[49]) and atmospheric $CO_2$ concentration and chlorophyll fluorescence (OCO-2/3[30], TROPOspheric Monitoring Instrument (TROPOMI)[50], and Geostationary Carbon Cycle Observatory (GeoCarb)[51]) are expected to help us better address these challenges.

**Methods**

**Annual AGB dataset during 2010-2019.** *In situ* measurements of forest AGB dynamics in the Amazon are limited to local forest inventory plots and seasonal direct biometric measurement plots[9, 10, 11]. Several studies have combined datasets from both forest inventory plots and remote sensing to generate spatial maps of forest AGB estimates at multi-year time frames[3, 12, 52], based on canopy height estimates from the Geoscience Laser Altimeter System (GLAS) LiDAR sampling strips and vegetation indices from optical images (MODIS). The recently developed L-VOD AGB dataset is one of major satellite-based data sources for monitoring inter-annual changes of AGB in the tropical regions[20, 21, 53, 54].

The L-VOD AGB data product was derived from the Soil Moisture and Ocean Salinity (SMOS) passive microwave satellite images L-VOD ascending product (version 1.6) developed by the French National Institute for Agricultural Research (INRAE) and Center for the Study of the Biosphere from Space (CESBIO)[21, 53, 54]. Our previous work by co-authors Fan et al.[20] used both ascending observations (acquired at 6am) and descending observations (acquired at 6pm) over the pan-tropic zone. L-VOD has diurnal dynamics because of leaf water content changes in a day. Here, we used the ascending observations, because at 6 am the water refilling process through plant xylems restores the leaf water potential to values close to the root-zone soil water potential, and an equilibrium is reached in the soil–plant–atmosphere continuum[55]. As a result, the ascending observations at 6am are less sensitive to the plant water stress than the descending observations at 6 pm and are more pertinent to monitor AGB[20]. The use of only ascending observations was possible in this study, as a lot of sub-daily observations were available over the Brazilian Amazon, which is an area which is very little impacted by noisy microwave interferences at L-band[20]. L-VOD also has seasonal dynamics as vegetation canopy changes over seasons. Several steps of data filtering were applied to retrieve relatively robust and stable annual estimates (mean and median) and all calculations are detailed in Fan et al.[20].

Here, we used the maximum L-VOD (L-VODmax, defined as 95% percentile each year), which occurs mostly within the wet season. During the wet season, the L-VODmax data are relatively independent of annual changes in the dielectric properties of vegetation, which may be assumed to be relatively constant from year to year. Note that we computed L-VOD changes for individual grid cells over years, and it is not our primary task to investigate spatial variations in these dielectric properties. In the long term, this may not be true, as there are changes in vegetation types. But over ten years, we can assume that over a given grid cell, average vegetation moisture content/dielectric properties during the wet period are about constant. We know that this is not a perfect assumption. However, this assumption was found to be quite well supported by the signatures of the intact forests (FAF > 99% each year) which have stable temporal L-VOD and L-VODmax at the selected site (Supplementary Fig. 3) and over the whole Brazilian Amazon (Supplementary Fig. 4). There are seasonal changes in L-VOD, but it recovers to the same value each year during the wet period, which suggests that changes in L-VODmax are only due to biomass changes and not due to changes in the dielectric properties.

As in Fan et al.[20], the SMOS L-VOD was converted to carbon density using previously published biomass maps[3, 12, 52] as references by regressions between annual median of L-VOD (2011) and AGB maps: annual median L-VOD values were converted into the unit carbon density (Mg C/ha) and then averaged. Here, we calculated two sets of L-VOD AGB products for each year using the equations (Eq. 1, Supplementary Table 2) generated based on L-VOD in 2011 and two biomass maps generated by Saatchi et al[3] and Baccini et al[12] of the tropical Americas and then averaged them to get annual AGB maps during 2010-2019. As for the L-VOD product, the L-VOD AGB dataset has a spatial resolution of ~25 km. Fan et al.[20] have done extensive spatial uncertainty analyses of AGB and AGB changes including (1) internal uncertainties associated with the L-VOD derived AGB estimates and (2) external uncertainties associated with different reference biomass maps and biomass stocks at continental scales. Combining internal and external errors, the relative spatial uncertainties associated with AGB and the AGB changes are on the order of 20-30% over the tropics and continents[20].

$$AGB = a \times \frac{\arctan(b \times (VOD - c)) - \arctan(-b \times c)}{\arctan(b \times (Inf - c)) - \arctan(-b \times c)} + d \quad (1)$$

where $a$, $b$, $c$ and $d$ are four best-fit parameters and $VOD$ is the yearly L-VOD data. The yearly L-VOD data calculated for 2011 was used in Eq. 1, as described by Rodriguez-Fernandez et al.[56], because 2011 was the first complete year after the SMOS commissioning phase.

The remote sensing datasets we used in our study provide temporally continuous changes in AGB and forest area, but all optical, active, and passive microwave images used to estimate AGB encounter various degrees of saturation where forest biomass is very high. However, the L-VOD AGB dataset saturates only at ~200 Mg C ha$^{-1}$ [22], which, according to Saatchi et al.[3] and Baccini et al.[12], only happens at 2.47% and 0.01%, respectively, of total pixels. Compared to previous studies that used high frequency VOD (LPRM, LPDR applied to AMSR-E/2)[57, 58], the L-VOD AGB dataset (version 2.0) shows a strong relationship between changes in AGB and changes in FAF (Fig. 2).

**Annual forest maps during 2010-2019.** We generated annual maps of forests in South America during 2007-2010 at 50-m spatial resolution, using the images from the ALOS PALSAR and time series data from the MOD13Q1 Terra Vegetation Indices data product at 16-day temporal resolution and 250m spatial resolution[24]. We use the FAO's forest definition in our forest mapping studies, that is, forest as a land parcel (0.5 ha or larger) covered by 10% or more tree cover with tree height > 5 meters at their maturity. The resultant annual PALSAR/MODIS forest map in 2010 have high accuracy (> 90%) using very high spatial resolution images and 2-m land cover maps[17, 24]. Here, we used the canopy height and canopy cover percentage datasets retrieved from the direct measurements of the three-dimension canopy structure from the GLAS observations onboard the NASA's ICESat-1[59] to assess the 50-m PALSAR/MODIS forest map in the Brazilian Amazon in 2010 in terms of FAO's forest definition. The derived ICESat-1 canopy cover percentage showed almost no bias when compared with airborne lidar estimates and was sensitive to signal dynamics over dense forests, even when canopy cover exceeded 80%. The ICESat-based canopy height and canopy cover percentage estimates were able to better characterize footprint-level canopy conditions than the existing products derived from conventional optical remote sensing[59]. There are 1.1 million ICESat-1 site observations in the Brazilian Amazon. We found that 98.5% of the PALSAR/MODIS forest pixels having canopy height > 5 meters and 94.4% of the PALSAR/MODIS forest pixels having canopy cover percentage > 10% (Supplementary Fig. 5). Overall, 93.8% of the PALSAR/MODIS forest pixels having canopy height > 5 meters and canopy cover percentage > 10% (Supplementary Figs. 5, 6).

We developed a pixel- and phenology-based algorithm to identify and map evergreen forests in individual years[1, 17]. The algorithm was based on the canopy phenology from analyses of time series Enhanced Vegetation Index (EVI) and Land Surface Water Index (LSWI) from the 8-day 500 m MOD09A1 data product[1, 17]. A unique physical feature of evergreen forests is that evergreen forests have green leaves throughout the year, which is well captured by time series EVI and LSWI data in a year. We applied the algorithm to time series MOD09A1 data over individual pixels in a year and generated annual maps of evergreen forests in the Brazilian Amazon from 2000 to 2019 in the cloud computing platform Google Earth Engine[17]. Limited by the data availability, the forest map for 2019 was generated based on MOD09A1 imagery from January 1st to December 3rd, 2019. We carried out a temporal consistency check procedure that uses a three-year moving window filter to remove the noise in individual pixels and increase temporal consistency of evergreen forest maps. We further calculated the annual gross loss and gain of forest area based on the forest map in 2001 after excluding all the pixels without cloud-free observations[17].

$$EVI = 2.5 \times \frac{\rho_{nir} - \rho_{red}}{\rho_{nir} + 6 \times \rho_{red} - 7.5 \times \rho_{blue} + 1} \quad (2)$$

$$LSWI = \frac{\rho_{nir} - \rho_{swir}}{\rho_{nir} + \rho_{swir}} \quad (3)$$

where $\rho_{blue}$, $\rho_{red}$, $\rho_{NIR}$, and $\rho_{SWIR}$ represent land surface reflectance values from MOD09A1 blue, red, Near Infrared (NIR), and Short-wave Infrared (SWIR) bands, respectively.

The evergreen forest maps had relatively high overall accuracy (~97%) in the Brazilian Amazon in 2000 and 2010 based on the extensive high spatial resolution ground reference maps[17]. The evergreen forest loss and gain also have relatively high accuracy based on 2,000 stratified random sample pixels. The overall accuracy of the evergreen forest loss and gain are 97.79% (± 0.64%) and 99.18% (± 0.27%), respectively. We aggregated the 50-m PALSAR/MODIS forest map into 500-m FAF map and compared the areas and spatial consistency between the evergreen forest maps and the PALSAR/MODIS forest maps in the Brazilian Amazon during 2007-2010. The evergreen forests and the PALSAR/MODIS forests reached over 98% consistency in the forest area and forest spatial distribution[17]. Annual maps of evergreen forests in the Brazilian Amazon during 2000-2017 were reported in a recent study[17] and we extended the dataset to 2019 in this study, using the same method. We also compared the 25-m PALSAR-based forest areas developed by the Japan Aerospace Exploration Agency (JAXA) and the 50-m PALSAR/MODIS forest areas with the MOD100 forest areas in the Brazilian Amazon during 2007-2010 and 2015-2017 (2017 is the newest forest data map). JAXA forest areas and MOD100 forest areas have good consistency (Supplementary Fig. 7). Therefore, PALSAR/MODIS forest and MOD100 forest maps can be used to analyze the forest area changes in the Brazilian Amazon. A comparison among the PALSAR/MODIS forest maps, the PRODES forest map and the GFW forest maps was already reported[17].

**Global Forest Watch (GFW) forest area dataset during 2010-2019**[15]**.** Tree cover is defined as vegetation higher than 5 meters. The GFW (version 1.7) product includes a tree cover map in 2000, annual tree cover gross loss in 2001-2019, and total tree cover gross gain in binary for 2001-2012 at a spatial resolution of 30 meters. The GFW products were generated from decision tree algorithms through the analysis of time series Landsat images acquired during the growing season. The GFW products of 2000-2012 interval were generated based on Landsat 7 thematic mapper plus (ETM+) images. The GFW products of 2011-2019 interval were generated based on Landsat 5 thematic mapper (TM), Landsat 7 ETM+, and Landsat 8 Operational Land Imager (OLI) images and updated methodology. Due to variation in mapping algorithms and date of content, tree cover, tree cover gross loss and gain cannot be compared accurately against each other. Comparisons between the original 2001-2010 data and the 2011-2019 update should be performed with caution. The GFW product was evaluated with an overall commission error of 13% and an overall omission error of 12%, though the accuracy varies by biome and thus may be higher or lower in any particular location. The data producers are 75% confident that the loss occurred within the stated year, and 97% confident that it occurred within a year before or after (https://www.globalforestwatch.org/map?map=eyJjZW50ZXIiOnsibGF0IjoyNywibG5nIjoxMn0sImJlYXJpbmciOjAsInBpdGNoIjowLCJ6b29tIjoyfQ%3D%3D&modalMeta=tree_cover_loss).

**PRODES forest area dataset (2010-2019)**[14]**.** The PRODES forest product was generated by the Brazilian National Institute for Space Research (INPE) to identify annual deforestation and forest area in the Brazilian Amazon. One or two Landsat images as cloud free as possible are used each year per location. The images are then masked to exclude non-forest and previous deforestation, using the previous year's analysis results. Finally, interpreters delineate deforested polygons (shapefile format) in the intact primary forests of the previous year. In this study, we used the annual deforestation area statistics in the Brazilian Amazon during 2010-2019 as reported by the INPE.

**Atmospheric $CO_2$ concentration dataset during 2015-2018.** We obtained daily column-averaged atmospheric $CO_2$ concentration data from the NASA Orbiting Carbon Observatory 2 (OCO-2)[60]. OCO-2 was launched into orbit on July 2, 2014 and flies in a near-polar orbit as part of the Afternoon Train (A-train) constellation of satellites, with a local overpass time of approximately 1:30pm. It has been recording spectra in the 0.76μm, 1.61μm, and 2.05μm spectral regions on a near-continuous basis for five years. The OCO-2 version 9 column-averaged atmospheric $CO_2$ concentration ($XCO_2$) dataset[61] during 2015-2018 is publicly available. Only observations with quality flag 0 (i.e. "good") were considered in the Amazon, Atlantic Ocean (10° S < latitude < 10°N & 60°W < longitude < 20°W), and Pacific Ocean (10°S < latitude < 10°N & 110°W < longitude < 85°W) at the same latitude (Supplementary Fig. 8), which avoids soundings with errors due to unscreened clouds and aerosols as well as low signal to noise ratio[61]. Individual soundings were aggregated to 1° by 1° along track to account for correlated errors between soundings that are close to one another in space and time, in line with the conclusions of Worden et al.[62].

**Active fire and burned area datasets during 2010-2019.** The annual active fire and burned area data in the Brazilian Amazon were calculated using the 8-day 1 km MOD14A2 (version 006)[63] and the monthly 500 m MCD64A1 (version 006)[64], respectively. Limited by the data availability, we used MOD14A2 and MCD64A1 acquired between January 1st to December 3rd and between January and October 2019. We first selected active fire observations with nominal and high confidence level and burned area observations with sufficiently valid data in the reflectance time series. We then generated annual active fire and burned area binary maps if active fire and burned area occurred in a year during 2010-2019, respectively.

**Annual precipitation dataset and Evapotranspiration (ET) during 2010-2019.** We calculated annual precipitation during 2010-2019 using observations from the Tropical Rainfall Measuring Mission (TRMM), a joint mission between NASA and the JAXA. We used the precipitation from the TRMM 34B2 product with a 3-hour temporal resolution and a 0.25° × 0.25° (latitude and longitude) spatial resolution[65]. We calculated the annual ET as the sum of the 8-day global terrestrial ET from MOD16A2 V105 product at 1km pixel resolution[66]. ET is the sum of evaporation and plant transpiration from the Earth's surface to the atmosphere. We then calculated the annual water deficit as a difference between annual total precipitation and ET for each year.

**Maximum cumulative water deficit and numbers of dry months.** The moist tropical canopy transpires about 100 mm per month according to the ground measurements in different locations and seasons in the Amazon[27]. Forest will be in a water deficit (WD) when monthly precipitation (P) is less than 100 mm per month. The annual maximum cumulative water deficit (MCWD) is the maximum value of the monthly accumulated WD per year, which is a useful indicator of meteorologically induced water stress[27]. We calculated the annual MCWD during 2010-2018 using the monthly precipitation of TRMM 3B43 at 0.25° spatial resolution. We also calculated the number of dry months with water deficit during 2010-2018 in the Brazilian Amazon.

$$\text{If } WD_{n-1(i,j)} - E_{(i,j)} + P_{n(i,j)} < 0;$$

$$\text{then } WD_{n(i,j)} = WD_{n-1(i,j)} - E_{(i,j)} + P_{n(i,j)}; \quad (4)$$

$$\text{else } WD_{n(i,j)} = 0$$

where $WD$, $E$, and $P$ are water deficit, evapotranspiration, and precipitation, respectively. $E$ is equal to 100 mm per month. $i$ and $j$ are the coordinates (column and row) for grid cells. $n$ is the number of months each year.

**Photosynthetically active radiation (PAR) dataset during 2010-2019.** We calculated the annual mean values in each year during 2010-2019 using the monthly PAR dataset from the National Centers for Environmental Prediction-Department of Energy (NCEP-DOE) Reanalysis-2, which has a spatial resolution of 1.875° × 1.905° (longitude and latitude). We then resampled the annual mean PAR values into ~25 km × 25 km grid cells using the near resampling approach.

**Contributions of deforestation and forest degradation to bulk aboveground biomass loss.** Forest degradation and deforestation are not two independent processes. Deforestation leads to forest degradation by creating edges and increasing the perimeter of forests exposed to sources of fire ignition, and degraded areas are more likely to be deforested. The gross AGB loss in a grid cell is controlled by gross forest area loss, forest degradation, and other mechanisms such as non-forest biomass density changes. For grid cells (~25 km × 25 km) with decreased tree cover fraction, if tree cover fraction is still larger than 10%, we attributed the AGB loss entirely to degradation. If the tree cover fraction is smaller than or equal to 10%, we attributed the AGB loss to deforestation. From these two end members, we attempted a simple estimate of deforestation versus degradation within each grid cell using the method proposed by ref.[40]. First, we calculate the gross bulk AGB loss in each 0.25° grid cell. Second, we multiply the gross forest area loss during 2011-2019 by the AGB density in 2010 to approximately estimate AGB loss from deforestation. Then we calculate the difference between gross AGB loss and this deforestation contribution as being from degradation.

$$\Delta AGB_{Gross\ loss} = f(\Delta AGB_{Gross\ forest\ area\ loss}, \Delta AGB_{Degradation}, Others) \quad (5)$$

$$\text{If } AGB_{t+1} - AGB_t < 0, \text{ then } \Delta AGB_{Gross\ loss} = \sum (AGB_{t+1} - AGB_t), 2010 \leq t \leq 2019. \quad (6)$$

$$\Delta AGB_{Gross\ forest\ area\ loss} \cong \sum (Gross\ forest\ area\ loss) \times AGB_{Density} \quad (7)$$

$$\Delta AGB_{Degradation} \cong \Delta AGB_{Gross\ loss} - \Delta AGB_{Gross\ foest\ area\ loss} \quad (8)$$

**Statistical analysis and spatial-temporal analysis.** The 500 m annual forest maps (500 m spatial resolution) were aggregated into ~25 km × 25 km grid cells in ArcGIS 10.1 to match the spatial resolution of the L-VOD AGB dataset. The total forest area (ha) and the FAF (%) were then calculated within each individual grid cells. To analyze the co-variations between annual AGB and FAF changes (Extended Data Fig. 4), six category layers were created based on the FAF map in 2010: 0%, (0, 20%], (20, 40%], (40, 60%], (60, 80%], and (80, 100%]. Then the anomaly values (Z-score) for the total forest area and the total AGB were calculated in each category during 2010-2019, respectively.

The linear relationship model analyses (two-tailed) and the relevant slope, $R^2$, and p values were calculated between annual AGB and FAF (%) within each grid cells in the Brazilian Amazon during 2010-2019 in Matlab R2017a. The linear regression slope and spatial $R^2$ values were calculated between annual FAF and AGB during 2010-2019 using R raster and maptools packages.

**Data Availability:** The annual evergreen forest maps[67] and AGB maps[68] are freely available in the GeoTIFF format at Figshare. The GFW product is available at http://earthenginepartners.appspot.com/science-2013-global-forest. The PRODES forest product is available at http://www.obt.inpe.br/OBT/assuntos/programas/amazonia/prodes. MOD14A2, MOD16A2, and MCD64A1 products are available at https://lpdaac.usgs.gov/data/. The TRMM product is available at https://pmm.nasa.gov/data-access/downloads/trmm. The PAR product is from the NCEP/DOE 2 Reanalysis data provided by the NOAA/OAR/ESRL PSD, Boulder, Colorado, USA, from their Web site at https://www.esrl.noaa.gov/psd/.

**Code Availability:** The code for evergreen forest mapping and spatial correlation analysis are freely available at Figshare[69]. The other data processing and analyses were mainly in ArcMap (https://desktop.arcgis.com/en/arcmap/).

**Reference cited in Methods**

**Correspondence statement:** Correspondence and requests for materials should be addressed to X.X. (xiangming.xiao@ou.edu) and J.-P.W. (jean-pierre.wigneron@inra.fr).

**Acknowledgements:** We thank the valuable comments and suggestions of Dr. Luiz Aragão and two other reviewers to improve this study. We thank the comments and discussion from Pierre Friedlingstein, Nicolas Vuichard, Dan Zhu, Markus Kautz, and Ben Poulter to the early version of this manuscript. Y.Q. and X.X. were supported by NASA Land Use and Land Cover Change program (NNX14AD78G), the Inter-American Institute for Global Change Research (IAI) (CRN3076), which is supported by the US National Science Foundation (GEO-1128040), and NSF EPSCoR project (IIA-1301789). Y.Q., X.X., S.C., X.W., R.D. and B.M. were supported by NASA 'Geostationary Carbon Cycle Observatory (GeoCarb) Mission' (GeoCarb Contract # 80LARC17C0001). J.P.W. was supported by the SMOS project of the TOSCA Programme from CNES, France (Centre National d'Etudes Spatiales). P.C. and S.S. were supported by RECCAP2 project which is part of the ESA Climate Change Initiative (contract no. 4000123002/18/I-NB) and the H2020 European Institute of Innovation and Technology (4C; grant no. 821003). S.S. was supported by the Newton Fund through the Met Office Climate Science for Service Partnership Brazil (CSSP Brazil). M.B. was supported by the European Research Council (ERC) under the European Union's Horizon 2020 research and innovation programme (grant agreement no. 947757 TOFDRY) and a DFF Sapere Aude grant (9064-00049B). L.F. was supported by the National Natural Science Foundation of China (Grant No. 41801247, 41830648), Natural Science Foundation of Jiangsu Province (Grant No. BK20180806). X.L. was supported by China Scholarship Council (201804910838). F.L. was supported by the Strategic Priority Research Program of the Chinese Academy of Sciences (XDA20010202).

**Authors' contribution:** X.X. and Y.Q. designed the overall study plan. Y.Q. and X.X. prepared the annual evergreen forest maps. J.-P.W, M.B., L.F., and X.L. prepared the annual L-VOD

AGB dataset. S.C. prepared the OCO-2 $XCO_2$ dataset. Y.Q., X.X., X.W., R.D., Y.Z., and F.L. carried our data processing and analysis. X.X., Y.Q., J.-P.W., P.C., M.B., S.S., and L.F. interpreted the results. Y.Q. and X.X. drafted the manuscript and all co-authors contributed to writing and revision of the manuscript.

**Competing interests:** The authors declare no competing interests.

**Figure legends**

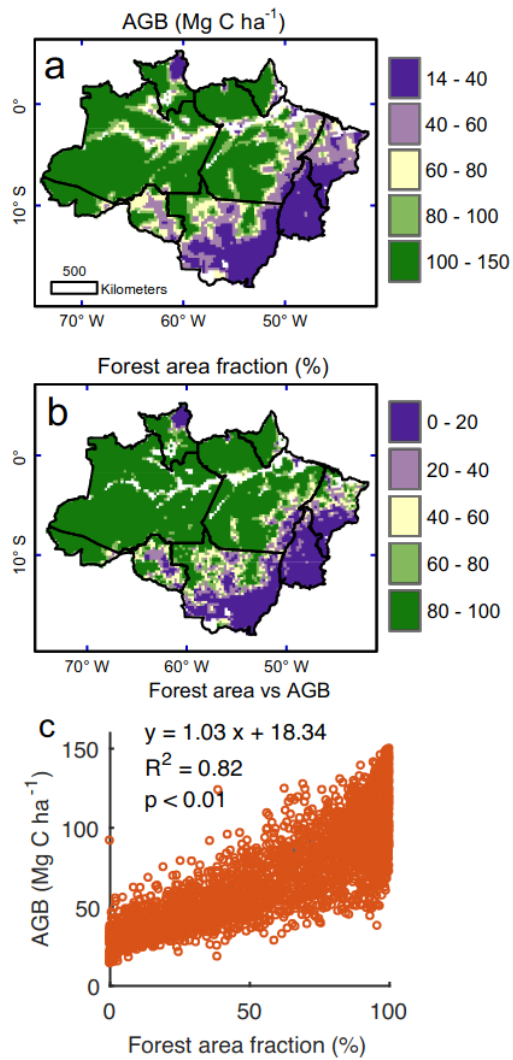

**Figure 1.** Spatial distributions of aboveground biomass (AGB) and forest area fraction (FAF, %) and their linear regression relationship within 0.25° (~25 × 25 km) grid cells. (a) Spatial distribution of averaged AGB (Mg C/ha) in 2019. (b) Spatial distribution of forest area fraction (%) in 2019. (c) Linear regression analysis between AGB and forest area fraction in 2019 (number of pixels = 5,656).

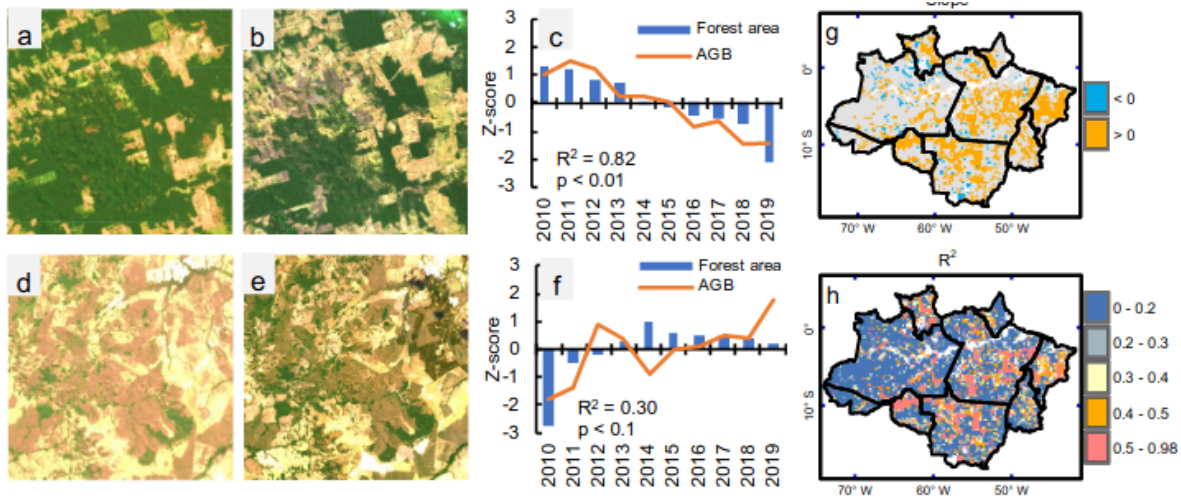

**Figure 2.** Inter-annual variation of forest area fraction and AGB during 2010-2019. (a) – (f) are two ~25×25 km grid cells with forest area loss (a and b, central Lat: 6.7S, Lon: 55.2W) and forest gain (d and e, central Lat:17.4S, Lon: 53.4W). (a) and (b) USGS/NASA Landsat images acquired on July 3th, 2010, and August 9th, 2019, and (c) annual anomaly values (Z-score) of forest area and AGB in the forest area loss grid cell (a and b). (d) and (e) Landsat images acquired on August 27th, 2010, and September 5th, 2019, respectively. (f) annual anomaly values (Z-score) of forest area and AGB in the forest area gain grid cell (d and e). (g) Map of the linear regression slope between annual forest area fraction and AGB during 2010-2019. The grey grid cells have temporal $R^2$ less than 0.3. (h) Map of temporal $R^2$ between annual forest area fraction and AGB during 2010-2019.

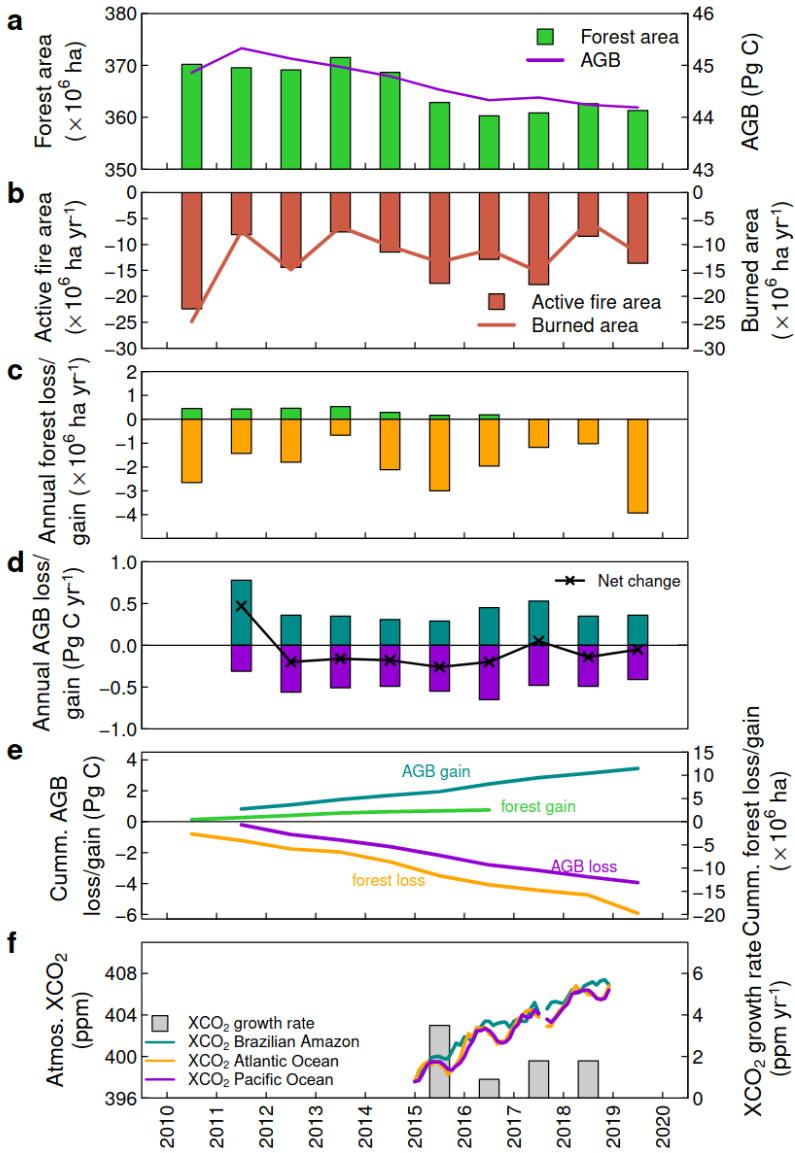

**Figure 3.** Inter-annual variations of annual AGB and forest areas in the Brazilian Amazon during 2010-2019. (a) Annual AGB estimated from L-VOD data and forest area estimated from MODIS data for each year. (b) Annual active fire area estimated by MOD14A2 and annual burned area estimated by MCD64A1. (c) Annual gross forest area loss and gain. (d) Inter-annual AGB changes (gross gain, gross loss, and net change). (e) Accumulated gross loss and gain of AGB and forest area. (f) Monthly atmospheric $XCO_2$ from OCO-2 observation in the Brazilian Amazon, Atlantic Ocean, and Pacific Ocean. To make the figure clear, we did not show the standard deviation values.

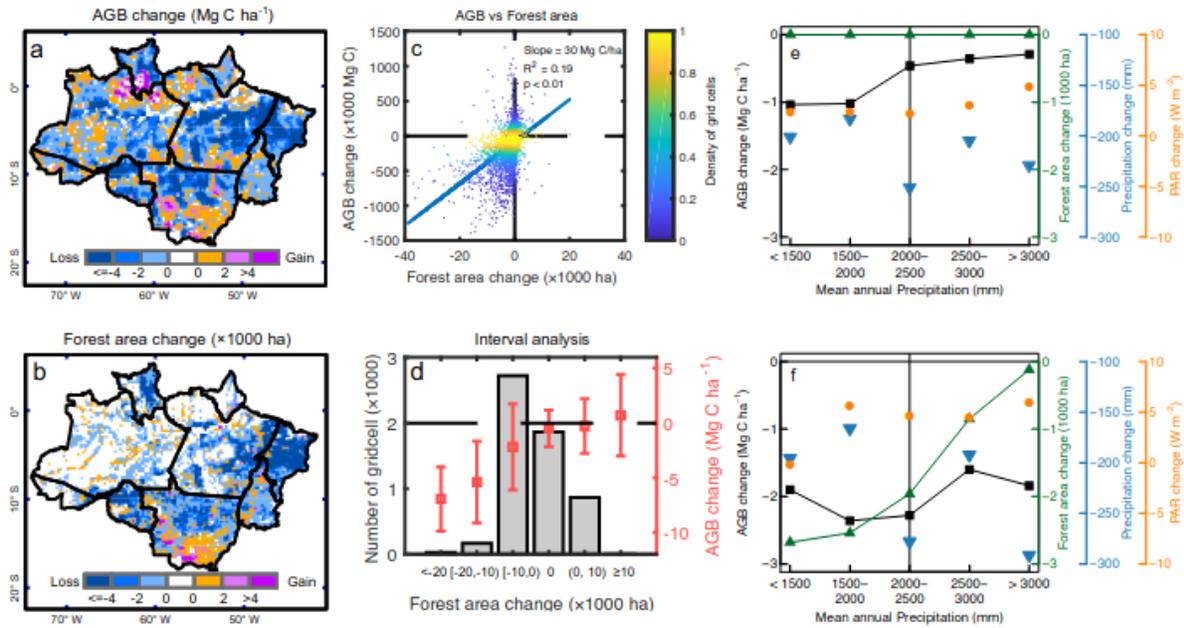

**Figure 4.** The changes of average AGB and forest area within 0.25˚ (~25×25 km) grid-cells before and after the 2015 extreme El Niño in 2010-2013 and 2015-2018. (a) Spatial distribution of the AGB change as a difference between the second and first periods. (b) Spatial distribution of forest area change. (c) Scatter plot and regression between forest area and AGB changes across grid-cells. (d) Grid cell numbers (gray bars) for different bins of forest area change and average values and standard deviation of AGB change. (e) AGB, forest area, precipitation, and photosynthetically active radiation (PAR) changes only over the only for grid-cells with no forest area change grouped into different bins of mean annual precipitation (500-mm intervals). (f) Same but for grid-cells with a forest area loss, of up to 10×10$^3$ ha between the two periods.

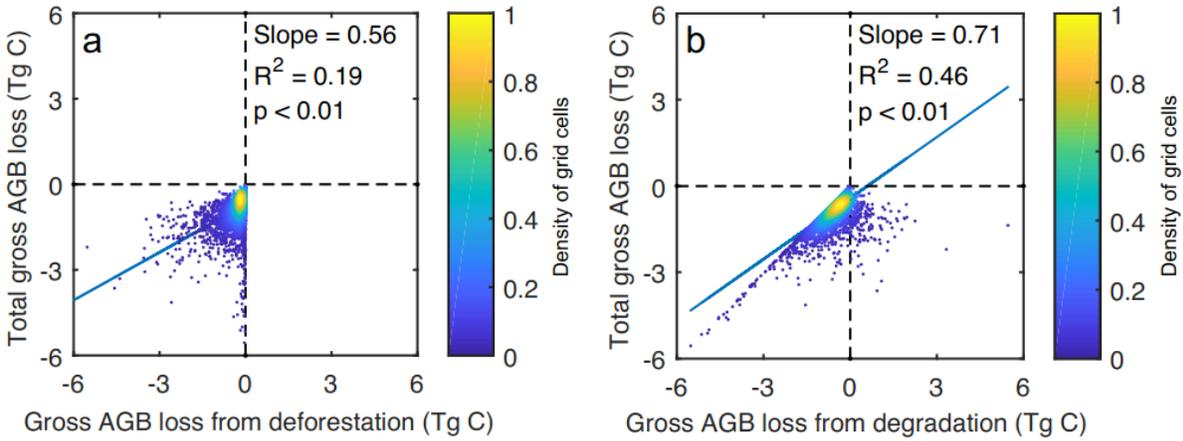

**Figure 5.** Total gross AGB loss from deforestation and forest degradation in those grid cells with forest area loss (n = 4,830) during 2010-2019 in Brazilian Amazon. (a) Linear relationship between gross AGB loss from deforestation and total gross AGB loss. (b) Linear relationship between gross AGB loss from degradation and total gross AGB loss. We define a forest degradation to include all the mechanisms (selective logging, forest fires, mortality from climatic disturbances like storms and drought) that do not result into full deforestation.